\documentclass[12pt]{article}
\usepackage{epsfig}

\begin{document}

\title{      Adaptive  Coding and Prediction  of Sources with Large and Infinite Alphabets
 \footnote {
 The first author was  supported by INTAS
     grant no. 00-738 and Russian Foundation for Basic Research under Grant no. 03-01-00495.
 } }
\author{ Boris Ryabko$ ^\dag$ and  Jaakko Astola $ ^\ddag$ \\ \\
$ ^\dag $ {\small Siberian State  University of Telecommunication}
\\ {\small and Computer Science, Novosibirsk, Russia} \\ $ ^\ddag$
{\small Tampere International Center for Signal Processing} \\
{\small at the Technical University of Tampere, Finland}
 }
\date{}
\maketitle

\begin{abstract}
The problem of predicting a sequence $x_1,x_2,\cdots$ generated by
a discrete source with unknown statistics is considered. Each
letter $x_{t+1}$ is predicted using the information on the word
$x_1x_2\cdots x_t$ only. This problem is of  great importance for
data compression, because of its use to estimate probability
distributions for PPM algorithms and other adaptive codes. On the
other hand, such prediction is a classical problem which has
received much attention. Its history can be traced back to
Laplace. We address the problem where the sequence is generated by
an i.i.d. source with  some large (or even infinite) alphabet. A
method is presented for which the redundancy of the code goes to 0
even for infinite alphabet.

\end{abstract}


\section{Introduction }

    The problem of prediction  can be traced back to Laplace
\cite{FE, WB}. Presently, the problem of prediction is
investigated by many researchers because of its practical
applications and importance for probability theory, information
theory, statistics and other theoretical sciences, see \cite{Al,
Ki, Mo, No, Sp,WB}.

Consider a source  with unknown statistics which generates
sequences $x_1x_2\cdots$ of letters from an alphabet
$A=\{a_1,\cdots,a_n\}$. The underlying true distribution, which is
unknown, is indicated by the letter $p$. Let the source generate a
message $x_1\ldots x_{t-1}x_t\ldots $, $x_i\in A$ for all $i$, and
let $ \nu^t(a)$ denote the count of letter $a$ occurring in the
word $x_1\ldots x_{t-1}x_t $. After the
 first $t$ letters $x_1,\ldots, x_{t-1},x_t$ have been processed
the following letter $ x_{t+1}$ needs to be predicted. By
definition, the prediction is the set of non-negative numbers
$p^\ast(a_1|x_1\cdots x_t),\cdots, p^\ast(a_n|x_1\cdots x_t)$
which are estimates of the unknown conditional probabilities
$p(a_1|x_1\cdots x_t), \cdots,p(a_n|x_1\cdots x_t)$, i.e.\ of the
probabilities $p(x_{t+1}=a_i| x_1\cdots x_t)$; $i=1,\cdots,n$.

Laplace suggested the following prediction
$
p^*_L(a|x_1\cdots x_t) = (\nu^t(a) +1 )/ (t+ |A | ), $ where
 $|A |$ is the number of letters
in the alphabet $A,\,$\cite{FE,WB}. (The problem which Laplace
considered  was to estimate the probability that the sun will rise
tomorrow, given that it has risen every day since the creation.
Using our terminology, we can say that Laplace estimated
$p(r|rr\cdots r)$ and $p(\bar r|rr\cdots r)$, where $\{r,\bar r\}$
is the alphabet (\lq\lq sun rises\rq\rq , \lq\lq sun does not
rise\rq\rq) and the length of $rr\cdots r$ is the number of days
since the creation.)

We will estimate the precision of the prediction by the
Kullback-Leibler divergence between the true distribution $p(.|x_1
... x_t)$ and its estimation $ p^*_\gamma(.|x_1 ... x_t),$ where
$\gamma$ denotes a prediction method.

In data compression the average divergence is called redundancy
and plays a key rule. Let us comment on the relation to data
compression in more detail. An encoder can construct a code with
codelength $ l^{\ast}(a|x_1\cdots x_t)\approx -\log
p^{\ast}(a|x_1\cdots x_t)$ for any letter $a\in A$ (the
approximation may be as accurate as you like, if the arithmetic
code is used, see \cite{Ri,Mof}). An ideal encoder would base
coding on the true distribution $p$ and not on the prediction
$p^{\ast}$. The difference in performance (the redundancy)
measured by the average code length is given by $$ \sum_{a\in
A}p(a|x_1\cdots x_t)(-\log p^\ast (a|x_1\cdots x_t)) -\sum_{a\in
A}p(a|x_1\cdots x_t)(-\log p(a|x_1\cdots x_t))$$ $$ =\sum_{a\in
A}p(a|x_1\cdots x_t)\log \frac{p(a|x_1\cdots x_t)} {p^\ast
(a|x_1\cdots x_t)}\,\;.\; $$ So, we can see that from a
mathematical point of view the prediction and adaptive coding are
identical and can, therefore, be investigated together. Note that
such a scheme of adaptive (or universal) coding was suggested in
\cite{CW}  and now is called as a PPM algorithm.

It is known that the redundancy of the Laplace predictor is upper
bounded by $(|A|-1)/(t+1)$ \cite{Ry2,RT}, if the predictor is
applied to an i.i.d. source. Krichevsky \cite{Kr1, Kr2}
investigated the problem of optimal minimax predictor for i.i.d.
sources and showed that, loosely speaking, the redundancy of the
optimal predictor is asymptotically equal to $(|A|-1)/(2t) +
o(1/t).$ We can see that, on the one hand, the precision of the
predictors essentially depends on the alphabet size $|A|$. On the
other hand, there are many applications, where the alphabet size
is unknown beforehand and can be upper bounded only. Moreover,
quite often such a bound is infinity. That is why the problems of
prediction and, especially, adaptive coding for large and infinite
alphabet sources have been a subject in literature before, see
\cite{E,O1,O2,M1,WB}.

In this paper we suggest a scheme of adaptive coding (and
prediction) for a case where a source generates letters from an
alphabet with unknown (and even infinite) size. This scheme can be
applied along with Laplace, Krichevsky and any other predictors.
If the suggested scheme is applied to $s-$ letter source and $s$
is unknown, the redundancy  is asymptotically the same as if the
predictor is applied to the $(s+1)-$ letter source and the
alphabet size $(s+1)$ is known beforehand.

When the suggested scheme is  applied to an infinite alphabet, the
redundancy of the code goes to 0, if, loosely speaking, the
original representation of the alphabet letters has a finite
average word length. It will be shown that, in fact, this
condition is necessary for existing of such predictors.

We mainly consider a case of prediction for i.i.d. sources,
because the predictors for i.i.d. sources are used as a "core" in
the PPM scheme and other practically used adaptive codes and the
performance of those codes  effectively depends on the redundancy
of the i.i.d. predictors. Besides, all results can be easily
extended to Markov sources, using well known methods, see, for
example., \cite{Kr1, RT}.

\section{ Definitions and Preliminaries }

    Consider an alphabet $A=\{a_1,\cdots,a_n\}$ with $n\geq2$ letters
and denote by $A^t$ the set of words $x_1\cdots x_t$ of length $t$
from $A$. Let $p$ be a source which generates letters from $A$.
Formally, $p$ is a probability distribution on the set of words of
infinite length or, more simply, $p=(p^t)_{t\geq1}$ is a
consistent set of probabilities over the sets $A^t\,;\;t\geq1$. By
$M_0(A)$ we denote the set of Bernoulli (or i.i.d.) sources over
$A$.

Denote by $D(\cdot\|\cdot)$ the Kullback-Leibler divergence and
consider a source $p$ and a method $\gamma$ of prediction. The
{\it{redundancy}}  is characterized by the divergence
\begin{equation}\label{r0}
r_{\gamma,p}(x_1\cdots x_t)=D\left(p(\cdot\vert x_1\cdots
x_t)\|p_\gamma^\ast (\cdot|x_1\cdots x_t)\right)$$
 $$ =\sum_{a\in
A}p(a|x_1\cdots x_t)\log\frac{p(a|x_1\cdots x_t)} {p_\gamma^\ast
(a|x_1\cdots x_t)}.
\end{equation}
(Here and below $\log \equiv \log_2$.)

For fixed $t$, $r_{\gamma,p}$ is a random variable, because $x_1,
x_2, \cdots, x_t$ are random variables. We  define the
{\it{average redundancy}}  at time $t$ by
\begin{equation}\label{r1}
r^t(p\|\gamma)=E\,\left(r_{\gamma,p}(\cdot)\right)=\nonumber \,
\sum_{x_1\cdots x_t\in A^t}p(x_1\cdots
x_t)\,\,r_{\gamma,p}(x_1\cdots x_t).
\end{equation}

Related to this quantity we define the {\it{maximum average
divergence}} (at time $t$) by \begin{equation}\label{r2}
r^t(M\|\gamma)=\sup_{p\in M}r^t(p\|\gamma), \end{equation} where
$M$ is a set of sources.
 If the Laplace predictor
 \begin{equation}\label{L}
p^*_L(a|x_1\cdots x_t) = (\nu^t(a) +1 )/ (t+ |A | )
\end{equation}  is applied to i.i.d. source,  its average redundancy ($r^t(p\|L)$) is
upper bounded by $(|A |-1) \log e / (t+1),$ i.e.
\begin{equation}\label{L2} r^t(M_0\|L)\leq
(|A | -1)\log e/ (t+1),\end{equation} see \cite{Ry2,RT}.(Here $e =
2.718...$ is the Euler number and, as before, $ \nu^t(a)$ denote
the count of letter $a$ occurring in the word $x_1\ldots
x_{t-1}x_t.) $ In \cite{Kr2} Krichevsky investigated the problem
of optimal predictor for the set of i.i.d. sources and showed that
for any predictor $\pi$ the maximal redundancy is asymptotically
lower bounded by $ (|A|-1 ) \log e / 2t :$ $$
\lim_{t\rightarrow\infty} \sup\: ( 2t\:\, r^t(M_0\|\pi))\,\:  \geq
(|A|-1 )\log e. $$ He has also suggested the predictor
\begin{equation}\label{K1}
p^*_{K_1}(a|x_1\cdots x_t) = (\nu^t(a) + \delta )/ (t+ \delta|A |
),
\end{equation} where $\delta= 0.50922...,$
and shown that it is asymptotically optimal:
\begin{equation}\label{K} \lim_{t\rightarrow\infty} \sup\: (
2t\:\, r^t(M_0\|K_1))\,\: = (|A|-1) \log e .\end{equation} As we
mentioned above, predictors for i.i.d. sources can be easily
extended to Markov sources (see, for example, \cite{Kr1, RT}) and
to the general stationary and ergodic sources, as it was suggested
in \cite{Ry0,Ry1}. But, it is worth noting that, as it is shown in
\cite{Ry1}, there exist such stationary and ergodic sources that
their divergence does not go to 0. More precisely, for any
predictor $\gamma$ there exists such a  stationary and ergodic
source $\tilde{p},$ that $ \lim_{t\rightarrow\infty} \sup\: (\:\,
r^t(\tilde{p}\|\gamma))\,\: \geq \log |A| ,$  (see
\cite{Ry1,Al,Mo}). But, on the other hand, it is shown in
\cite{Ry0,Ry1} that there exists such a predictor $\rho$, that the
following average $ R^t(p\|\rho)= t^{-1}\: \sum_{i=1}^t
r^i(p\|\rho) $ goes to 0 for any stationary and ergodic source
$p,$ where  $t$ goes to infinity. We will focus our attention on
the per letter redundancy $r^t(\,\|\,)),$ but all estimates can be
easily extended  to the $R^t(\,\|\,).$

\section{ The new scheme}

Let, as before, $p$ be an i.i.d. source generating letters from
the alphabet $A$. The probability distribution $p(a), a \in A $ is
unknown and each letter $x_{t+1}$ should be predicted (or encoded)
using information on the word $x_1x_2\cdots x_t$ only.

The suggested scheme can be applied to any predictor (or
letterwise  coding), but we will use the Laplace predictor as the
main example. We start the description of the new scheme using a
simple example. Let $A = \{ a_0, a_1, a_2, \}$ and $t=4,\;
x_1x_2x_3x_4 = a_0 a_2 a_0 a_0 . $ The "common" Laplace predictor
is as follows: $$ p_L^*(x_5 = a_0) = (3+1)/(4 + 3)= 4/7, p_L^*(x_5
= a_1) = (0+1)/(4 + 3) = 1/7, $$ $$ p_L^*(x_5 = a_0 = (1+1)/(4 +
3)= 2/7, $$ see (\ref{L}). In this example we suggest to group
letters into two subsets $ A_0 = \{a_0, a_1 \}, A_1 = \{a_3 \} $
and carry out the prediction into two steps. First, the original
sequence $a_0 a_2 a_0 a_0 $ is represented as $ A_0 A_1 A_0 A_0 $
and belonging to the subsets is predicted as follows $$ p_L^*(x_5
\in A_0) = (3+1)/ (4+2) = 2/3, p_L^*(x_5 \in A_1)  = (1+1)/ (4+2)
= 1/3. $$ We know that $A_1$ contains one letter ($a_3$), hence,
$p_L^*(x_5 = a_3) = 1/3$. The sequence $ A_0 A_1 A_0 A_0 ,$ which
contains three letters $A_0$, is used for predicting  conditional
probabilities $p(x_{5}= a_i / x_5 \in A_0),\: i= 0,1 .$ Again, we
apply the Laplace predictor (\ref{L}) to the sequence $ A_0  A_0
A_0 = a_0 a_0 a_0 $ and obtain the following prediction: $p(x_{5}=
a_0 / x_5 \in A_0) = (3+1)/ (3+2) = 4/5, p(x_{5}= a_1 / x_5 \in
A_0) \\ = (0+1)/ (3+2) = 1/5 .$ So, combining all predictions, we
obtain $ p_L^*(x_5 = a_0) = (2/3 ) (4/5) = 8/15, $ $ p_L^*(x_5 =
a_1) = (2/3)(1/5) = 2/15, $ $
 p_L^*(x_5 = a_0) = 1/3.$ We can see that this prediction and the
"common" one are different.

It will be convenient to describe the general case using the
notation of a tree. Let $\Upsilon$ be a rooted tree, which
contains $|A|$ leaves, and let each leaf be marked by one letter
from $A$ in such a way that different leaves are marked by
different letters.

\begin{figure}[h]\label{fig1}
 \begin{center}
 \setlength{\epsfxsize}{2in}
 \centerline{\epsfbox{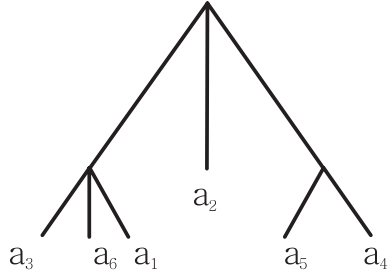}}
 \caption{$A_{\alpha_1}=\{a_1,a_3,a_6 \},
 A_{\alpha_{2}}=\{a_2\},A_{\alpha_{3}}=\{a_4,a_5\},A_{\alpha_{1,1}}=\{ a_3\},... \: $.}
 \end{center}
 \end{figure}
 We will mark each vertex $\mu \in \Upsilon $ by
a subset of $A_\mu$ as follows. We consider the subtree
$\Upsilon_\mu$ whose root is the vertex $\mu$ and define the
subset $A_\mu$ by the set of all letters, which mark the leaves of
the subtree $\Upsilon_\mu$. Note that $\Upsilon_{root} = A.$ We
denote  vertexes of the depth one by $\alpha_i, i = 1,...,k $, the
vertexes of the depth 2 by $\alpha_{i,j},  j = 1, ..., k_i, i =
1,...,k,$ etc., where $k$ is the number of the depth 1 vertexes,
$k_i$ is the number of the sons of the vertex $i$, etc. The
prediction is carried out as follows. First a generated sequence $
x_1 \ldots x_t $ is represented as the sequence $A_{\alpha_{i_1}}
A_{ \alpha_{i_2}}\ldots A_{\alpha_{i_t}}$, where $\alpha_{i_j}$ is
such a vertex of the depth one,  that the letter $x_j$ belongs to
the subset $A_{\alpha_{i_j}}.$ Then, $A_{\alpha_{i_1}} A_{
\alpha_{i_2}}\ldots A_{\alpha_{i_t}}$ is considered as the
sequence generated by an i.i.d. source with the alphabet $ \{
A_{\alpha_i}, i = 1,...,k \} $ and the next letter
$A_{\alpha_{i_{t+1}}}$ is predicted (say, by Laplace predictor).
In fact,  $\, p^*_L( A_{\alpha_i}) $ is the estimation of $\,Pr\,(
x_{t+1} \in A_{\alpha_i}).$ Then, for each  vertex $\alpha_{i,j},
i = 1,...,k, j = 1, ..., k_i$ of the depth 2, which is not a leaf,
we estimate the probability $Pr( x_{t+1} \in A_{\alpha_{i,j}} /
x_{t+1} \in A_{\alpha_i})$. For this purpose for each $i$ we
consider all letters $A_{\alpha_i}$ in the sequence
$A_{\alpha_{i_1}} A_{ \alpha_{i_2}}\ldots A_{\alpha_{i_t}}$ and
organize the following sequence of their sons $A_{\alpha_{i,j_1}}
... A_{\alpha_{i,j_s}}, $ whose length $s$ equals the count of
occurrences $A_{\alpha_i}$ in the sequence $A_{\alpha_{i_1}} A_{
\alpha_{i_2}}\ldots A_{\alpha_{i_t}}.$ This sequence is considered
as a generated by an i.i.d. source and the next letter
$A_{\alpha_{i,j_s}}$ is predicted. As a result, we obtain  $\,
p^*_L( A_{\alpha_{i,j}}) $, which are, in fact, the estimations of
the probabilities $Pr( x_{t+1} \in A_{\alpha_{i,j}} / x_{t+1} \in
A_{\alpha_i})$. And so on. Then we calculate the predictor
$P^*_\Upsilon (x_{t+1} = a )$ as a product $\,Pr\,( x_{t+1} \in
A_{\alpha_i})\:$ $Pr( x_{t+1} \in A_{\alpha_{i,j}} / x_{t+1} \in
A_{\alpha_i})$  $Pr( x_{t+1} \in A_{\alpha_{i,j,m}} / x_{t+1} \in
A_{\alpha_{i,j}})$ $...$, where $a \in A_{\alpha_{i}},$ $a \in
A_{\alpha_{i,j}},$ $a \in A_{\alpha_{i,j,m}}, ...\: .$

The following example is to illustrate all steps. Let the alphabet
$A$ be $\{ a_1, \ldots, a_6 \}$ and the tree $\Upsilon$ is shown
on the figure 1. Let the generated sequence $x_1 \ldots x_t$ be
$a_3 a_1 a_5 a_5 a_2 a_5 a_4 a_2 a_3.$ According to the tree
$\Upsilon,$ firstly this sequence is represented as $A_{\alpha_1}
A_{\alpha_1} A_{\alpha_3} A_{\alpha_3} A_{\alpha_2} A_{\alpha_3}
A_{\alpha_3} A_{\alpha_2} A_{\alpha_1}$ and the next letter
$x_{t+1}$ is predicted as follows: $ P^*_{L_\Upsilon}( x_{t+1} \in
A_{\alpha_1} ) = (3+1) /(9+3),$ $ P^*_{L_\Upsilon} ( x_{t+1} \in
A_{\alpha_2} ) = (2+1) /(9+3),$ $ P^*_{L_\Upsilon} ( x_{t+1} \in
A_{\alpha_3} ) = (4+1) /(9+3).$ Then we divide the  sequence of
subsets $A_{\alpha_1} A_{\alpha_1} A_{\alpha_3} A_{\alpha_3}
A_{\alpha_2} A_{\alpha_3} A_{\alpha_3}\\ A_{\alpha_2}
A_{\alpha_1}$ into the three following subsequences
$A_{\alpha_1}A_{\alpha_1}A_{\alpha_1}$, $A_{\alpha_2}A_{\alpha_2}$
and $A_{\alpha_3}A_{\alpha_3}A_{\alpha_3}A_{\alpha_3}.$ The set
$A_{\alpha_2}$ contains only one letter $a_2,$ hence, the
prediction of this letter coincides with the prediction of the
subset $A_{\alpha_2}$: $p^*_{L_\Upsilon}( x_{t+1} = a_2) $ $ =
P^*_{L_\Upsilon} ( x_{t+1} \in A_{\alpha_2} ) = (2+1) /(9+3).$ The
first subsequence $A_{\alpha_1}A_{\alpha_1}A_{\alpha_1}$ is
represented as $a_3 a_1 a_3$ and the next letter is predicted by $
P^*_{L_\Upsilon}( x_{t+1} = a_1 / x_{t+1} \in A_{\alpha_1}) = (1+
1)/ (3+3),$ $ P^*_{L_\Upsilon}( x_{t+1} = a_3 / x_{t+1} \in
A_{\alpha_1}) = (2+ 1)/ (3+3),$ $ P^*_{L_\Upsilon}( x_{t+1} = a_6
/ x_{t+1} \in A_{\alpha_1}) = (0+ 1)/ (3+3).$ The third
subsequence $A_{\alpha_3}A_{\alpha_3}A_{\alpha_3}A_{\alpha_3}$ is
represented as $a_5 a_5 a_5 a_4$ and we obtain the following
prediction: $ P^*_{L_\Upsilon}( x_{t+1} = a_4 / x_{t+1} \in
A_{\alpha_3}) = (1+ 1)/ (4+2),$ $ P^*_{L_\Upsilon}( x_{t+1} = a_5
/ x_{t+1} \in A_{\alpha_3}) = (3+ 1)/ (4+2).$ Finally, we obtain
$$ P^*_{L_\Upsilon}( x_{t+1} = a_1) = (4/12) (2/6),
P^*_{L_\Upsilon}( x_{t+1} = a_2) = (3/12),$$ $$  P^*_{L_\Upsilon}(
x_{t+1} = a_3) = (4/12) (3/6),  P^*_{L_\Upsilon}( x_{t+1} = a_4) =
(5/12) (2/6),$$ $$  P^*_{L_\Upsilon}( x_{t+1} = a_5) = (5/12)
(4/6), P^*_{L_\Upsilon}( x_{t+1} = a_6) = (4/12) (1/6).$$ The
properties of the suggested scheme are as follows.

\textbf{Theorem 1.} Let $ x_1 \ldots x_t $ be a sequence generated
by an i.i.d. source from an alphabet $A$. If the letter $x_{t+1}$
is predicted by the suggested scheme according to a tree
$\Upsilon$ with the set of vertex $\Lambda$ , then the following
upper bound for  the precision (or redundancy) (\ref{r1}) is
valid:
\begin{equation}\label{t1}
r^t(p\| L_\Upsilon ) \leq \,\log e \: \sum_{\lambda \in\, \Lambda
} \min \{\, (\sigma(\lambda) -1)/ (t+1) ,\: p(A_\lambda) \}\, ,
\end{equation}
where $\sigma(\lambda)$ is the number of sons of the vertex
$\lambda,$ $p(A_\lambda) = \sum_{a \in A_\lambda} \: p(a),$
$L_\Upsilon$ is a notation of the predictor.

{\it Proof} is in Appendix.

\textbf{ Comment}. The summands, which correspond to leaves in
(\ref{t1}), are equal to 0, because $A_\lambda =1$, if $\lambda$
is a leaf. So, if we define the set of internal vertex by
$\Lambda_{int}$ we can rewrite (\ref{t1}) as follows: $$ r^t(p\|
L_\Upsilon ) \leq\, \log e\:\sum_{\lambda \in\, \Lambda_{int} }
\min \{\,
 (\sigma(\lambda) -1)/ (t+1) ,\: p(A_\lambda) \}\, \,. $$

\textbf{ Corollary. } The "common" Laplace predictor corresponds
to the tree which consists of  a root with $|A|$  sons (leaves).
From Theorem 1 we immediately obtain the  upper bound (\ref{L2}).

\section{The unknown and infinite alphabet}

Of course, the suggested scheme can be generalized in such a way
that the tree $\Upsilon, $ which determines the algorithm of
prediction, depends on the encoded part $x_1 ... x_t.$ 
We consider one such scheme, which is close, in spirit, to some
methods basing on the estimate of an escape probability
\cite{M1,WB}, and show how Theorem 1 can be used for obtaining an
asymptotic  estimations of the redundancy in the case where the
algorithm of prediction (or the tree $\Upsilon$) depends on the
sequence.

In order to describe this predictor we define the subset $A_0^t$,
which contains all 0-frequent letters at the moment $t$  and let
$A^t_+ = A- A_0^t.$ The predictor $\eta_t$ is defined  as follows:

$$ \eta^t(a)=\cases{(\nu^t(a)+1)/(t+ |A^t_+|+1)  ,&if $\nu^t(a)
> 0$;\cr
           1/((t+ |A^t_+|+1)\, |A^t_0|),&if $\nu^t(a) = 0$ .}
$$ In words, the letter $x_{t+1}$ is encoded according to  the
 tree $\Upsilon_t,$ whose root has $|A^t_+| +1$ sons. Of these,
$|A^t_+|$  are leaves corresponding to the letters from the set
$A^t_+$
 (i.e. their frequencies are greater than 0). One son of the root
 corresponds to the set $A_0^t$ and has $|A_0^t|$ sons, which, in
 turn, correspond to the letters from $A_0^t$. This code is close to
some methods, which are based on estimate of the escape
 probability, see \cite{WB}. It is interesting that,
asymptotically, the increase of the redundancy is the same as if
the alphabet size would  increase by one. In words, if the
suggested scheme is applied to $s-$ letter source and $s$ is
unknown, the redundancy is the same as if the predictor is applied
to $(s+1)-$ letter source and the alphabet size $(s+1)$ is known
beforehand. More precisely, the following property is valid.

\textbf{Theorem 2.} $\:$ Let the predictor  $\eta_t$ be applied to
an i.i.d. source $p$ with an alphabet $A$. Then, the following
inequality is valid: $$ \limsup_{t\rightarrow\infty}\,(\,t\,\,
r^t(p)\|\eta_t)\, )\, \leq\: \min \{s, |A|-1 \},  $$ where $s$ is
the number of the alphabet letters whose probability is not 0.

{\it Proof} is in Appendix.

\textbf{Comment.} The Krichevsky predictor (\ref{K1}) can be used
along with the described scheme (instead of the Laplace
predictor). If we define this predictor as $ \tilde{\eta}_t,$ then
 the following inequality is valid: $$
\limsup_{t\rightarrow\infty}\,(\,2 \,t\,\,
r^t(p)\|\tilde{\eta}_t)\, )\, \leq\: \min \{s, |A|-1 \}.  $$

If we compare these inequalities with the redundancy of the common
Laplace predictor (\ref{L2}) and the Krichevsky one (\ref{K}), we
can see that the "payment" for the lack of knowledge of the
alphabet size is asymptotically as large as it could be caused
increasing
 the alphabet by 1 letter. In words, if the letters with nonzero
probabilities are known beforehand, the redundancy of the common
Laplace predictor is $(s-1)/ (t+1),$ where $s$ is the number of
letters with nonzero probabilities.

 Now we consider the case of infinite  alphabet $A = \{ a_1, ...,
 a_n,...\}
\:$. We suppose that there is  an i.i.d. source, which generates
letters from $A$ with a probability distribution $p.$ Obviously,
the first letter $x_1$ must be encoded by an encoder and decoded
by a decoder based on an initial code (over some finite alphabet),
which is known to both the encoder and the decoder. (Otherwise,
the first letter cannot be transmitted.) We denote this initial
code by $c = \{ c_1, c_2, ... \}$ and suppose that the letter
$a_i$ is encoded by the word $c_i$. Of course, we also suppose
that the code $c$ is uniquely decodable. Moreover, we will suppose
that $c$ is the prefix code, because it is known that for each
decodable code $c$ there exists such a prefix code $c^*$ that
their codeword lengths are equal ($|c_i| = |c^*_i|$ for all $i$),
see \cite{Ga}.

 If one wants to use this code for encoding and decoding the first
letter $x_1,$ the average codeword length of this letter should be
finite, i.e.
\begin{equation}\label{c1}
E_p(c_i)\, \equiv \,\sum_{i=1}^{\infty} p(a_i)\, | c_i | \: <\:
\infty\,.
\end{equation}
For example, if the (binary) code and the probability distribution
are as follows: $$ c_1 =0, c_2 = 10, c_3= 110, c_4= 1110, ...; \:
p(a_1)=1/2, p(a_2)= 1/4, p(a_3)= 1/8, ..., $$ then the average
codeword length $E_p(|c_i|) = 2$ bits and, hence, such a code can
be used for a coding $x_1.$ On the other hand,  consider the same
probability distribution and a new code $\bar{c}$ such that
$|\bar{c}_i| = 2^{i}.$ This code cannot be used, because its
average codeword length is infinite and the first letter $x_1$
cannot be transmitted from an encoder to an decoder. So, our
requirement is as follows: We do not know the source
probabilities, but know  a code $c,$ whose  average codeword
length is finite. The following theorem shows that this
requirement is sufficient for existing such an adaptive code,
whose redundancy goes to 0. In a certain sense  this requirement
is necessary, because nobody can transmit the first letter of the
generated sequence, if he/she does not know such a code.

\textbf{Theorem 3.} Suppose that there are a source  generated
letters from an infinite alphabet $A$ with (unknown) distribution
$p$ and a prefix code $c(a), a \in A$  over some finite alphabet
such that the average length $E_p(|c(a)|)$ is finite. Denote by
$\Gamma$ the tree which corresponds to the code $c$ and let
$L_\Gamma$ be the Laplace predictor corresponding to the tree
$\Gamma.$ Then the redundancy of this method goes to 0: $$ \lim_{t
\rightarrow\infty} r^t(p\| L_\Gamma ) = 0. $$

{\it Proof} is in Appendix.

\textbf{Corollary.} If the source alphabet $A$ is infinite, but it
is known that only a finite number of letters have nonzero
probabilities, the average codeword length (\ref{c1}) is finite
for any code and, hence, Theorem 3 is valid for any prefix code.

\section{Appendix}
{\it The proof of the theorem 1.} We will use  two following
lemmas.

\textbf{ Lemma 1. } Let a Bernoulli source  generate $t$ letters
from the  alphabet $A^* = \{A, \bar{A} \}$ with probabilities
$p(A)$ and $p(\bar{A})$ and let $\vartheta(A)$ be a count of
occurrence of the letter $A$ in the generated sequence. Then the
following inequality is valid: $$\, p(A) \,\, E_p (\: 1/
(\vartheta(A) + 1)  \leq \min \{p(A), \:1/(t+1) \, \}, $$ where
$E_p$ is an expectation.

{\it Proof of the lemma 1.} The inequality $ p(A) \: E_\vartheta
(\: 1/ (\vartheta(A) + 1)  \leq p(A)$ is followed from the obvious
inequality $E_p (\: 1/ (\vartheta(A) + 1) \leq E_\vartheta (\: 1)
= 1.$

The second inequality  $\, p(A) \,\, E_\vartheta (\: 1/
(\vartheta(A) + 1) \leq  \:1/(t+1) \, $ is proved as follows: $$
p(A) \,\, E_\vartheta (\: 1/ (\vartheta(A) + 1) = p(A)\,
\sum_{j=0}^{t} \bigl(
\begin{array}{c} t
\\j\end{array} \bigr) p(A)^j(1-p(A))^{t+1-j}\,)$$

$$ \leq   \frac{1}{t+1} \sum_{s=0}^{t+1} \bigl(
\begin{array}{c} t+1
\\s\end{array} \bigr) p(A)^s(1-p(A))^{t+1-s}\,) =\frac{1}{t+1}. $$
The lemma is proven.

\textbf{ Lemma 2. } Let a Bernoulli source  generate $t$ letters
from the finite alphabet $A = \{a \}$ with probabilities $p(a), a
\in A .$ Then the redundancy of the "common" Laplace predictor
(\ref{L}) can be upper bounded by $\log e \frac{|A|-1}{t+1}.$
(Though Lemma 2 is known for the Laplace predictor (\cite{Ry2,RT})
we give the details of the proof for the convenience of the
reader.)

 We
employ the general inequality $$ D(\mu\|\eta)\leq \log e \:
(-1+\sum_{a\in A}\mu(a)^2/ \eta(a)), $$
 valid for any distributions $\mu$ and $\eta$ over $A$ (follows from
the elementary inequality $\ln x\leq x-1)$ From the definition of
the redundancy (or precision) (\ref{r0}), (\ref{r1}) and (\ref{L})
we obtain
 $$
r^t(p\|L)= E_{p^t} D(p(\cdot\mid x_1\cdots x_t)\|p^\ast (\cdot\mid
x_1\cdots x_t))$$ $$=E_{p^t} (D(p\|p_L^\ast (\cdot\mid x_1\cdots
x_t))
\leq \log e\, (-1+\sum_{x_1\cdots x_t\in A^t}p(x_1\cdots x_t)$$
$$\sum_{a\in A}\frac {p(a)^2(t+|A|)}{\nu_a(x_1\cdots x_t)+1}\,)
=\log e ( -1+\sum_{a\in A}\sum_{i=0}^t\frac
{p(a)^2(t+|A|)}{i+1}\,$$ $$ ( \bigl(\begin {array}{c} t\\
i\end{array}\bigr) p(a)^i(1-p(a))^{t-i} )) = \log e (
-1+\frac{t+|A|}{t+1}\sum_{a\in A}p(a) $$ $$ \sum_{i=0}^t \bigl(
\begin{array}{c} t+1\\ i+1\end{array}\bigr)
p(a)^{i+1}(1-p(a))^{t-i} \,)
\leq \log e (-1+\frac{t+|A|}{t+1}\sum_{a\in A}p(a)$$
$$\sum_{j=0}^{t+1} \bigl( \begin{array}{c} t+1
\\j\end{array} \bigr) p(a)^j(1-p(a))^{t+1-j}\,) =\log e \, \frac{|A|-1}{t+1}. $$
The lemma is proven.

Consider a letter $a \in A$ and let $\alpha_{i_1}$,
$\alpha_{i_1,i_2}$, $\ldots, $ $\alpha_{i_1,i_2,...,i_s}$ be a
sequence of vertexes such that $$ A_{\alpha_{i_1}} \supset
A_{\alpha_{i_1,i_2}} \ldots \supset A_{\alpha_{i_1,i_2,...,i_s}}
$$ and $A_{\alpha_{i_1,i_2,...,i_s}} = \{a\}.$ The estimation of
the probability (or prediction) $p(x_{t+1}= a)$ can be represented
as follows $$p^*_{L_\Upsilon}(a) = \left((\nu^t(A_{\alpha_{i_1}})+
1)/ (t+ \sigma( root))\right)\;\; \left((\nu^t(A_{\alpha_{{i_1,
i_2}}})+ 1)/ (\nu^t(A_{\alpha_{i_1}})+ \sigma(\alpha_{i_1}))
\right)\; $$ $$(\nu^t(A_{\alpha_{{i_1, i_2, i_3}}})+
1)/(\nu^t(A_{\alpha_{{i_1,i_2}}})+ \sigma(\alpha_{{i_1,i_2}})) \;
\ldots (\nu^t(A_{\alpha_{{i_1, ..., i_s}}})+ 1)/(
\nu^t(A_{\alpha_{i_1, ..., i_{s-1}}})+ \sigma(\alpha_{i_1, ...,
i_{s-1}} )),$$ where $\nu^t(A_\alpha)$ is a number of occurrences
of letters from the subset $A_\alpha$ in the sequence  $ x_1
\ldots x_t. $ From this equality and the definition of the
redundancy we obtain the following equality: $$ r^t(p|L_\Upsilon)=
E_{p(x_1 \ldots x_t)} ( \sum_{a \in A} \, p(a) \log
(p(a)/p^*_{L_\Upsilon}(a))   =\: \sum_{a \in A} \, p(a) $$ $$\;
\{\; \log \frac{p(A_{\alpha_{i_1}})} { ((\nu^t(A_{\alpha_{i_1}})+
1)/ (t+ \sigma( root)}   $$ $$ + \: \log \frac{
(p(A_{\alpha_{i_1,i_2}})/p(A_{\alpha_{i_1}}))} {
(\nu^t(A_{\alpha_{{i_1, i_2}}})+ 1)/ (\nu^t(A_{\alpha_{i_1}}) +
\sigma(\alpha_{i_1}))} $$

$$ + \: \log \frac{
(p(A_{\alpha_{i_1,i_2,i_3}})/p(A_{\alpha_{i_1,i_2}}))} {
((\nu^t(A_{\alpha_{{i_1, i_2,i_3}}})+ 1)/
(\nu^t(A_{\alpha_{i_1,i_2}} + \sigma(\alpha_{i_1,i_2})))}\; ... $$

$$ + \: \log \frac{
(p(A_{\alpha_{i_1,...,i_s}})/p(A_{\alpha_{i_1,...,i_{s-1}}})) } {
((\nu^t(A_{\alpha_{{i_1,...,i_s}}})+ 1)/
(\nu^t(A_{\alpha_{i_1,...,i_{s-1}}}) +
\sigma(\alpha_{i_1,...,i_{s-1}})) )}\; \}\; ) . $$ Grouping
summands, we obtain

$$ r^t(p|L_\Upsilon)= E ( \; \;\sum_{i_1}\, p(A_{\alpha_{i_1}})\,
\log \frac{p(A_{\alpha_{i_1}})} { (\nu^t(A_{\alpha_{i_1}})+ 1)/
(t+ \sigma( root))} )  $$

$$+ \sum_{i_1} p(A_{\alpha_{i_1}}) E( \sum_{i_2}\, \frac{
p(A_{\alpha_{i_1,i_2}})}{p(A_{\alpha_{i_1}})} \log
\frac{p(A_{\alpha_{i_1,i_2}})/ p(A_{\alpha_{i_1}}) } {
((\nu^t(A_{\alpha_{i_1,i_2}})+ 1)/ (\nu^t(A_{\alpha_{i_1}})+
\sigma( \alpha_{i_1})})) +\; ...  $$

$$+ \sum_{i_1} p(A_{\alpha_{i_1}}) \sum_{i_2}\, \frac{
p(A_{\alpha_{i_1,i_2}})}{p(A_{\alpha_{i_1}})} ...E( \sum_{i_s}\,
\frac{
p(A_{\alpha_{i_1,...,i_s}})}{p(A_{\alpha_{i_1,...,i_{s-1}}})} $$
$$ \log \frac{p(A_{\alpha_{i_1,...,i_s}})/
p(A_{\alpha_{i_1,...,i_{s-1}}}) } {
(\nu^t(A_{\alpha_{i_1,...,i_s}})+ 1)/ (\nu^t(A_{\alpha_{i_1,...,
i_{s-1}}})+ \sigma( \alpha_{i_1,..., i_{s-1}})})) + ..., $$ where
$E(\:)$ means an expectation. It is worth noting that such a
grouping is correct for the case of infinite $A,$ because all
values $E(\:)$ are nonnegative.

Having taken into account this equality and Lemma 2 we can upper
bound the redundancy as follows:

$$ r^t(p|L_\Upsilon) \leq \log e \; \{ \; ( \sigma (root) -1)/ (
t+1) ) $$

$$+ E \, (\, \sum_{i_1} p(A_{\alpha_{i_1}}) \frac{\sigma(
\alpha_{i_1})-1 } {
 (\nu^t(A_{\alpha_{i_1}})+
1)}) +\; ...$$

$$+ \sum_{i_1} p(A_{\alpha_{i_1}}) \sum_{i_2}\, \frac{
p(A_{\alpha_{i_1,i_2}})}{p(A_{\alpha_{i_1}})} ... E \,
(\sum_{i_{s-1}}\, \frac{
p(A_{\alpha_{i_1,...,i_{s-1}}})}{p(A_{\alpha_{i_1,...,i_{s-2}}})}
$$ $$ \frac{\sigma( \alpha_{i_1,..., i_{s-1}} )-1 } {
(\nu^t(A_{\alpha_{i_1,..., i_{s-1}}})+1  )}\:+ ...\} .$$

Let us estimate one summand, say, $$\sum_{i_1} p(A_{\alpha_{i_1}})
\sum_{i_2}\, \frac{ p(A_{\alpha_{i_1,i_2}})}{p(A_{\alpha_{i_1}})}
  \,
\sum_{i_{3}}\,E( \frac{
p(A_{\alpha_{i_1,i_2,i_{3}}})}{p(A_{\alpha_{i_1,i_{2}}})} \:
\frac{\sigma( \alpha_{i_1,i_2, i_{3}} )-1 } {
(\nu^t(A_{\alpha_{i_1,i_2, i_{3}}})+1  )} .$$ (General case is
obvious).
From Lemma 1 we obtain the following inequalities

$$\sum_{i_1} p(A_{\alpha_{i_1}}) \sum_{i_2}\, \frac{
p(A_{\alpha_{i_1,i_2}})}{p(A_{\alpha_{i_1}})}
 E \,
(\sum_{i_{3}}\, \frac{
p(A_{\alpha_{i_1,i_2,i_{3}}})}{p(A_{\alpha_{i_1,i_{2}}})}
 \frac{\sigma( \alpha_{i_1,i_2, i_{3}} )-1 } {
(\nu^t(A_{\alpha_{i_1,i_2, i_{3}}})+1  )})$$ $$= \sum_{i_1}
p(A_{\alpha_{i_1}}) \sum_{i_2}\, \frac{
p(A_{\alpha_{i_1,i_2}})}{p(A_{\alpha_{i_1}})} \sum_{i_{3}}\,
(\sigma( \alpha_{i_1,i_2, i_{3}} )-1) E \, (\frac{
p(A_{\alpha_{i_1,i_2,i_{3}}})}{p(A_{\alpha_{i_1,i_{2}}})}
 \frac{1} {
(\nu^t(A_{\alpha_{i_1,i_2, i_{3}}})+1  )})$$

$$\leq \sum_{i_1} p(A_{\alpha_{i_1}}) \sum_{i_2}\, \frac{
p(A_{\alpha_{i_1,i_2}})}{p(A_{\alpha_{i_1}})} \sum_{i_{3}}\,
(\sigma( \alpha_{i_1,i_2, i_{3}} )-1) E \min \{\frac{
p(A_{\alpha_{i_1,i_2,i_{3}}})}{p(A_{\alpha_{i_1,i_{2}}})},
\frac{1}{(\nu^t(A_{\alpha_{i_1,i_2}})+1  )} \} $$

$$= \sum_{i_1} p(A_{\alpha_{i_1}}) \sum_{i_2}\,  \sum_{i_{3}}\,
(\sigma( \alpha_{i_1,i_2, i_{3}})-1)  \min \{\frac{
p(A_{\alpha_{i_1,i_2,i_{3}}})}{p(A_{\alpha_{i_1}})}, E(
\frac{p(A_{\alpha_{i_1,i_2}})/ p(A_{\alpha_{i_1}})} {
(\nu^t(A_{\alpha_{i_1,i_2}})+1 )})  \} $$

$$\leq \sum_{i_1} p(A_{\alpha_{i_1}}) \sum_{i_2}\,  \sum_{i_{3}}\,
(\sigma( \alpha_{i_1,i_2, i_{3}} )-1)  \min \{\frac{
p(A_{\alpha_{i_1,i_2,i_{3}}})}{p(A_{\alpha_{i_1}})},E( \frac{1} {
(\nu^t(A_{\alpha_{i_1}})+1 )} ) \} $$

$$= \sum_{i_1}  \sum_{i_2}\,  \sum_{i_{3}}\, (\sigma(
\alpha_{i_1,i_2, i_{3}} )-1)  \min \{
p(A_{\alpha_{i_1,i_2,i_{3}}}),E( \frac{p(A_{\alpha_{i_1}})} {
(\nu^t(A_{\alpha_{i_1}})+1 )} ) \} $$

$$\leq \sum_{i_1}  \sum_{i_2}\,  \sum_{i_{3}}\, (\sigma(
\alpha_{i_1,i_2, i_{3}} )-1)  \min \{
p(A_{\alpha_{i_1,i_2,i_{3}}}), \frac{1} { (t+1 )}  \} .$$ From
this estimation and the last upper bound for $ r^t(p|L_\Upsilon)$
we obtain the inequality $$  r^t(p|L_\Upsilon) \leq
 \log e \; \{ \; ( \sigma (root) -1)/ (
t+1) )  +\sum_{i_1}  \, (\sigma( \alpha_{i_1} )-1) \min \{
p(A_{\alpha_{i_1}}), \frac{1} { (t+1 )}  \} $$ $$+ \sum_{i_1}
\sum_{i_2}\,  (\sigma( \alpha_{i_1,i_2} )-1)  \min \{
p(A_{\alpha_{i_1,i_2}}), \frac{1} { (t+1 )}  \} $$ $$+ \sum_{i_1}
\sum_{i_2}\,  \sum_{i_{3}}\, (\sigma( \alpha_{i_1,i_2, i_{3}} )-1)
\min \{ p(A_{\alpha_{i_1,i_2,i_{3}}}), \frac{1} { (t+1 )} \}+ ...
\} .$$ In fact, the last inequality is (\ref{t1}) and Theorem 1 is
proven.

{\it The proof of the theorem 3.} Let $\varepsilon > 0$ be some
number. There exists such a finite subset $A_\varepsilon \subset
A$ that
\begin{equation}\label{ap1}
\sum_{a \in (A - A_\varepsilon ) } p(a) |C(a)| < \varepsilon/(2
\log e) ,
\end{equation}
because the sum $\sum_{a \in A  } p(a) |C(a)|$ is finite. The
redundancy of the code $L_\Gamma$ can be represented as two
following summands:
\begin{equation}\label{ap2}
r^t(p \|L_\Gamma) = \sum_{a \in A_\varepsilon } p(a) \log ( p(a)/
p^*_{L_\Gamma^t}(a )) + \sum_{a \in (A-A_\varepsilon) } p(a) \log
( p(a)/ p^*_{L_\Gamma^t}(a )),
\end{equation} where $L^t_\Gamma$ is the Laplace probability
assignment at the moment $t$. The number of vertex in the subtree
corresponding to $A_\varepsilon$ is finite. Hence, if we  apply
the theorem 1, we obtain the following upper bound for the first
summand in (\ref{ap2}): $$ \sum_{a \in A_\varepsilon } p(a) \log (
p(a)/ p^*_{L_\Gamma^t}(a ))  \leq const / (t+1). $$ Hence, there
exists such $t_0$ that the first summand in (\ref{ap2}) is less
than $\varepsilon/2,$  if $t
> t_0 .$ Let us estimate the second summand in (\ref{ap2}).
First we define the subset $\Lambda_\varepsilon,$ which contains
all vertex belonged to passes from the root to each letter from
$A- A_\varepsilon$. Applying the theorem 1, we obtain

$$ \sum_{a \in (A-A_\varepsilon) } p(a) \log ( p(a)/
p^*_{L_\Gamma^t}(a )) \leq \log e \: \sum_{\lambda \in
\Lambda_\varepsilon} p(A_\lambda) = \log e \: \sum_{\lambda \in
\Lambda_\varepsilon} (\,\sum_{ a \in A_\lambda } \:p(a)\, ). $$ If
we define the number of vertex belonged to the passe  from the
root to the leaf $a$ as $\chi(a)$ we can rewrite the latest
inequality as follows.

$$ \sum_{a \in (A-A_\varepsilon) } p(a) \log ( p(a)/
p^*_{L_\Gamma^t}(a )) \leq \log e \: \sum_{a \in A_\varepsilon}
p(a) \chi(a)  . $$ By definition the codeword length $|C(a)|$ of
any letter $a$ is equal to $\chi(a)$. Hence, having taken into
account this fact,  from the latest inequality and (\ref{ap1}) we
obtain the following upper bound for the second summand: $$
\sum_{a \in (A-A_\varepsilon) } p(a) \log ( p(a)/
p^*_{L_\Gamma^t}(a )) \leq \varepsilon/2 .$$ So, each summands in
(\ref{ap1}) is upper bounded by $\varepsilon/2$ and it is true for
any positive $\varepsilon.$ Theorem 2 is proven.


\end{document}